\documentclass{pjay}
\usepackage{caption,floatrow}
\usepackage{upgreek}
\usepackage{combelow}
\usepackage{comment}
\usepackage{cite}
\usepackage{tikz,graphicx}
\usepackage{amsmath}
\usepackage{color}
\usepackage[mathscr]{eucal}
\usepackage{amssymb}

\def\XXint#1#2#3{{\setbox0=\hbox{$#1{#2#3}{\int}$}
     \vcenter{\hbox{$#2#3$}}\kern-.5\wd0}}

\newcommand{\be}[1]{\begin{equation} \label{#1} }
\newcommand{\bea}[1]{\begin{eqnarray} \label{#1} }

\newcommand{\bfi}{\begin{figure}}
\newcommand{\efi}{\end{figure}} 
\newcommand{\ee}{\end{equation}}
\newcommand{\eea}{\end{eqnarray}}
\newcommand{\bib}{\bibitem}
\newcommand{\lbl}{\label}

\newcommand{\w}{{\omega}}

\newcommand{\ve}{{\bf e}}

\newcommand{\vb}{{\bf b}}

\newcommand{\vg}{{\bf g}}

\newcommand{\vh}{{\bf h}}
\newcommand{\vj}{{\bf j}}

\newcommand{\vf}{{\bf f}}
\newcommand{\vp}{{\bf p}}
\newcommand{\vm}{{\bf m}}
\newcommand{\vrp}{{\bf r}}

\newcommand{\vd}{{\bf d}}

\newcommand{\vzh}{{\bf \hat{z}}}

\newcommand{\vr}{{\bf r}}
\newcommand{\vH}{{\bf H}}
\newcommand{\vE}{{\bf E}}
\newcommand{\vM}{{\bf M}}
\newcommand{\vP}{{\bf P}}

\newcommand{\vB}{{\bf B}}
\newcommand{\vF}{{\bf F}}
\newcommand{\vD}{{\bf D}}
\newcommand{\vJ}{{\bf J}}
\newcommand{\vG}{{\bf G}}

\newcommand{\dT}{\overline{\bf T}}
\newcommand{\dI}{\overline{\bf I}}

\newcommand{\dcT}{\mbox{\boldmath ${\overline{\mathcal T}}$\unboldmath}}

\newcommand{\vnh}{{\bf \hat{n}}}

\newcommand{\vphih}{{\bf \hat{\mbox{\boldmath {$\phi$}}}}}

\newcommand{\eps}{\epsilon}

\def\e{\begin{equation}}
\def\f{\end{equation}}
\def\_#1{{\bf #1}}

\def\.{\cdot}

\begin{document}
\setcounter{page}{1}
\pjheader{}
%

\title[]{\vspace{-18mm}{\large\bf Electromagnetic Force and Momentum in Classical Macroscopic Dipolar Media\\[2mm]{\small Arthur D. Yaghjian}}\\{\small\rm Electromagnetics Research, 115 Wright Road, Concord, MA 01742 USA  (a.yaghjian@comcast.net)}}
%
%
%
%
{\small
\begin{abstract}
{\bf Using realistic classical models of microscopic electric-charge electric dipoles {\color{black}and electric-current (Amperian) magnetic dipoles, it is proven that the Einstein-Laub macroscopic electromagnetic force on a macroscopic-continuum volume of these classical dipoles} equals the sum of the microscopic electromagnetic forces on the discrete classical dipoles in that volume.  The internal (hidden) momentum of the discrete Amperian magnetic dipoles is rigorously derived and properly included in the determination of the macroscopic force from the spatial averaging of the microscopic forces. Consequently, the Abraham/Einstein-Laub rather than the Minkowski macroscopic electromagnetic-field momentum density gives the total microscopic elec\-tro\-mag\-netic-field momentum in that volume.  The kinetic momentum is found for the volume of the macroscopic continuum from Newton's relativistic equation of motion.   It is shown that the difference between the kinetic and canonical {\color{black}momenta in a volume of the macroscopic continuum is equal to the sum of the ``hidden electromagnetic momenta''  within the electric-current magnetic dipoles and within hypothetical magnetic-current electric dipoles} replacing the electric-charge electric dipoles in the classical macroscopic continuum.  To obtain the correct unambiguous value of the force on a volume inside the continuum from the force-momentum expression, it is mandatory that the surface of that volume be hypothetically separated from the rest of the continuum by a thin free-space shell.  Two definitive experiments performed in the past with time varying fields and forces are shown to conclusively confirm the Einstein-Laub/Abraham formulation over the Minkowski formulation.}
\end{abstract}
%
%
%
\
\section{Introduction}\label{Intro}
Although the determination of the detailed fields and polarizations of atoms and molecules requires quantum physics, most bulk materials below optical (or even higher) frequencies are accurately described by the classical Maxwell macroscopic equations for dipolar continua \cite[sec. 77]{L&L}.  Moreover, the microscopic (molecular) dipoles producing the macroscopic dipolarization can be adequately modeled pragmatically by classical electric-charge electric dipoles and Amperian (circulating-electric-current) magnetic dipoles, irrespective of their actual quantum origin. Indeed, most of the widely used physics and engineering textbooks in electromagnetics, such as \cite{L&L,Stratton,P&P,Jackson,VanBladel}, confine themselves predominantly to classical electromagnetic theory with classical models of electric and magnetic dipoles. 
\par
Nonetheless, since Maxwell published his electromagnetic equations, it has remained uncertain as to how to correctly determine the time varying classical macroscopic force on a volume of dipolar material subject to time varying electromagnetic fields.  In particular, uncertainty continues as to whether to use the Abraham definition \cite{Abraham1,Abraham2} of macroscopic electromagnetic-field momentum (which is also contained in the Einstein-Laub \cite{E&L} macroscopic force-momentum equation) or the Minkowski definition \cite{Minkowski} of macroscopic electromagnetic-field momentum (or some other definition of macroscopic electromagnetic-field momentum), each of which leads to a different instantaneous time-domain macroscopic electromagnetic force (even when their predicted time-averaged macroscopic electromagnetic {\color{black}forces are} the same) \cite{Griffiths,McDonald,Mansuripur2017,Silveirinha}.
\par  
A major obstacle preventing the determination of the correct macroscopic electromagnetic force and electromagnetic-field momentum in polarized material approximated by a classical macroscopic continuum has been the uncertainty of the relationship between the forces on individual dipoles and the forces on a distribution of these dipoles composing the macroscopic continuum.  A second major obstacle has been the absence of a definitive determination of the electromagnetic time-domain force on classical microscopic Amperian magnetic dipoles, which approximate the magnetic dipoles found in nature.  In this paper, these obstacles are overcome and it is determined that the correct macroscopic electromagnetic force and electromagnetic-field momentum for classical macroscopic dipolar continua are given by the Einstein-Laub macroscopic force and the Abraham/Einstein-Laub (rather than Minkowski -- or any other) macroscopic electromagnetic-field momentum.  An underlying important requirement of all the derivations is that the surface of any volume of the dipolar material under consideration lie in a hypothetical thin free-space shell separating the volume from the rest of the material, so that the volume contains a discrete number of dipoles.
\section{Microscopic Force and Momentum}
We assume that we are dealing with a macroscopic dipolar continuum (solid or fluid) whose molecules or inclusions have electric and magnetic dipole moments that can be modeled electromagnetically by {classical} microscopic electric charge and electric current with fields that obey the following Maxwell differential equations in SI (mksA) units
\begin{subequations}
\label{1}
\be{1a}
\nabla\times\ve(\vrp,t) +\frac{\partial \vb(\vrp,t)}{\partial t} =0
\ee
\be{1b}
\frac{1}{\mu_0}\nabla\times\vb(\vrp,t) -\eps_0\frac{\partial \ve(\vrp,t)}{\partial t} = \vj(\vrp,t)
\ee
\be{1c}
\nabla\cdot\vb(\vrp,t) =0
\ee
\be{1d}
\eps_0\nabla\cdot\ve(\vrp,t) = \varrho(\vrp,t)
\ee
\end{subequations}
where $\ve(\vrp,t)$ and $\vb(\vrp,t)$ are the primary microscopic electric and magnetic fields at the position $\vrp$ and time $t$, scalar $\varrho(\vrp,t)$ and vector $\vj(\vrp,t)$ are the microscopic electric-charge and electric-current densities, and $\eps_0$ and $\mu_0$ are the free-space permittivity and permeability ($1/\sqrt{\mu_0\eps_0}= c$, the free-space speed of light).  Note that since there are no polarization densities in (\ref{1}), it follows that the microscopic electric displacement vector $\vd$ (secondary electric field) is given by $\vd =\eps_0\ve$ and the microscopic secondary magnetic field $\vh$ is given by $\vh =\vb/\mu_0$.  The charge and current densities, $\varrho(\vrp,t)$ and $\vj(\vrp,t)$, can be considered continuous functions of $\vrp$, even though they can form charges and currents of discrete dipoles occupying indefinitely small regions of space.
\par
With the help of these microscopic Maxwell equations, the microscopic Lorentz-force density
\be{fd}
\vf(\vrp,t)=\varrho(\vrp,t)\ve(\vrp,t)+\vj(\vrp,t)\times\vb(\vrp,t)
\ee
can be shown to satisfy the equation \cite[sec. 2.5]{Stratton}
\be{2}
\vf(\vrp,t) + \eps_0\frac{\partial}{\partial t}[\ve(\vrp,t)\times\vb(\vrp,t)] = \nabla\cdot\dcT(\vrp,t)
\ee
where the microscopic electromagnetic stress dyadic $\dcT(\vrp,t)$ is defined with the help of the unit dyadic $\dI$ as
\be{3}
\dcT = \eps_0\left(\ve\ve -\frac{1}{2}\dI|\ve|^2\right) + \frac{1}{\mu_0}\left(\vb\vb -\frac{1}{2}\dI|\vb|^2\right).
\ee
If the Lorentz force is written as the time rate of change of a ``Lorentz momentum'' density $\vg_L(\vrp,t)$, that is, $\vf(\vrp,t) = \partial\vg_L(\vrp,t)/\partial t$, and $\vg_f(\vrp,t)=\eps_0[\ve(\vrp,t)\times\vb(\vrp,t)]$ designates the microscopic {\color{black}``electromagnetic-field momentum''} density, then (\ref{2}) can be rewritten as
\be{4}
\nabla\cdot\dcT(\vrp,t) = \frac{\partial}{\partial t}[\vg_L(\vrp,t)+\vg_f(\vrp,t)]
\ee
which shows that $\nabla\cdot\dcT(\vrp,t)$ is the time rate of change of the total microscopic electromagnetic momentum density.
\par
Integrating (\ref{4}) over a volume $V$ with surface $S$ that encloses all the charge-current, and assuming (as a thought experiment) that all the charge is held fixed so there is no current (just electrostatic charge and field) until the charge is released at $t=0$, we have $\vg_f(\vrp,t\le0)=0$ and $\int_V \vf(\vrp,t\le0)dV=0$ (since the charge and field are electrostatic for $t\le0$).  Then integrating over time from $0$ to $t$ yields
\be{5}
\int\limits_0^t\int\limits_V\nabla\cdot\dcT(\vrp,t')dV dt' =  \int\limits_0^t\int\limits_S\vnh\cdot\dcT(\vrp,t')dS dt' = \int\limits_V [\vg_L(\vrp,t)+\vg_f(\vrp,t)] dV 
\ee
where $\vnh$ is the unit normal to $S$ pointing out of $V$ and $\int_V\vg_L(\vrp,t)dV = \int_0^t\int_V \vf(\vrp,t')dVdt'$, which equals $0$ for $t\le0$.  This equation shows that $\vnh\cdot\dcT(\vrp,t)$ represents the total electromagnetic momentum flow in the $-\vnh$ direction since $\vnh$ points out of $V$.  We see that if the surface $S$ is far enough away that the radiated fields have not had time to reach $S$ in the time $t$,  then the surface integral in (\ref{5}) is zero, that is, the total microscopic electromagnetic momentum in $V$ is zero and
\be{6}
\int\limits_V [\vg_L(\vrp,t)+\vg_f(\vrp,t)] dV = 0
\ee
for all time $t$ as long as no radiation has crossed $S$.
Since $t$ can take on any value as long as the surface $S$ of $V$ is chosen large enough, and both $\int_V\vg_L(\vrp,t\le0)dV=0$ and $\vg_f(\vrp,t\le0)=0$, equation (\ref{6}) expresses the conservation of total microscopic electromagnetic momentum in $V$ and confirms that indeed the microscopic electromagnetic-field momentum density $\vg_f(\vrp,t)$ can be treated as a legitimate physical momentum (because $\vg_L(\vrp,t)$ is a physical electromagnetic-force-produced momentum).
\par
So far, nothing has been said about the kinetic momentum of the charge carriers.  Certainly, the microscopic electromagnetic force density $\vf(\vrp,t)$ will, in general, change the kinetic momentum and energy of the charge carriers in $V$ but this does not affect the validity of the purely electromagnetic-momentum relationships in (\ref{2})--(\ref{6}).  Kinetic momentum {\color{black}as well as ``canonical momentum'' is} introduced in Section \ref{Canon}, where the two momenta are shown to be related by ``hidden electromagnetic momenta.''  It is assumed throughout that the macroscopic electromagnetic fields and momenta of the thermal motion of the molecules are either negligible or lie outside of the  bandwidth of the applied and induced macroscopic electromagnetic fields and momenta.
\section{Macroscopic Force and Momentum}\lbl{Macro}
The ambiguity in the macroscopic force on a volume of a dipolar material can be demonstrated directly from Maxwell's homogeneous (no macroscopic free charge $\rho$ and current $\vJ$, only polarization) equations for a macroscopic dipolar continuum \cite[ch. 1]{Stratton}
\begin{subequations}
\lbl{M1}
\be{M1a}
\nabla\times\vE + \frac{\partial \vB}{\partial t} = 0
\ee
\be{M1b}
\nabla\times\vH - \frac{\partial \vD}{\partial t} = 0
\ee
\be{M1c}
\nabla\cdot\vB = 0
\ee
\be{M1d}
\nabla\cdot\vD = 0
\ee
\end{subequations}
with the constitutive relations
\be{M2a}
\vD = \eps_0\vE +\vP,\;\;\;\;\vB = \mu_0(\vH +\vM).
\ee
\footnote{}\footnote{}\footnote{\vspace{-10mm}}
The vectors $\vP$ and $\vM$ are the macroscopic electric polarization and the magnetic polarization (magnetization) densities and the $(\vr,t)$ dependence of all the fields and polarizations have been suppressed.  The term ``macroscopic'' refers to fields and sources obtained by spatially averaging the microscopic fields and sources at each instant of time over electrically small volume elements $\Delta V$ that (in the medium) contain many discrete (isolated to an indefinitely small region of space) dipoles.  The term ``dipolar continuum'', which can be solid or fluid, simply means that the medium obeys the Maxwell dipolar equations in (\ref{M1},\ref{M2a}).  Thus, the combined term ``macroscopic dipolar continuum'' refers to a medium composed of discrete dipoles that, upon spatial averaging, obeys, to a good approximation, the Maxwell dipolar equations in (\ref{M1},\ref{M2a}).\footnote{\lbl{foot1}Contrary to what is sometimes stated in the historical literature, Maxwell (and not the ``Maxwellians'') determined all the equations in (\ref{M1}) \cite{Maxwell,Yaghjian-Reflections,Yaghjian-MaxwellPolarization} for the \textit{mathematically defined} fields of an ideal dipolar continuum where the polarization densities are continuous functions of position throughout the medium rather than composed of discrete dipoles as in a macroscopic dipolar continuum \cite{Yaghjian_PowerEnergy}.  It is unequivocally shown in \cite{Yaghjian_PowerEnergy}, \cite[sec. 2.1.10]{H&Y}  that Maxwell's equations in (\ref{M1}) for the mathematically defined fields of an ideal dipolar continuum also apply (approximately) to macroscopic dipolar media if and only if the surfaces $\Delta S$ of the defining macroscopic volumes $\Delta V$ lie in free space and do not intersect the discrete dipoles.  It follows that the force and momentum expressions obtained from (\ref{M1}) for a volume $V$ with a surface $S$ that lies within the polarization densities $\vP$ or $\vM$ are unambiguously defined if and only if the surface $S$ is placed within the free space of a hypothetical thin shell that separates $V$ from the rest of the continuum (created by removing the polarization densities within the shell without changing the adjacent polarization densities), so that the total bound charge and current densities in every $V$ are zero, that is, there are a discrete number of electric and magnetic dipoles in $V$  \cite{Yaghjian_PowerEnergy}, \cite[secs. 2.1.1 and 2.1.10]{H&Y}.  For an ideal Maxwellian continuum, the thin free-space shells ensure that any delta functions in the polarization charge and current densities at the surface of the volume $V$ are taken into account.   This requirement for unambiguous field and force-momentum expressions, namely that the surfaces of the volumes do not cut through the dipoles, is also stated by Einstein and Laub \cite{E&L}, Landau and Lifshitz \cite[secs. 6 and 29]{L&L}, and De Groot and Suttorp \cite[pp. 195--196]{D&S}.  \textit{These hypothetical thin free-space shells containing $S$ are assumed throughout the present paper and are crucial to a consistent formulation and determination of dipolar electromagnetic force and momentum}.} 
The same macroscopic Maxwell equations in (\ref{M1},\ref{M2a}) can be derived, for example, using electric-current (Amperian) magnetic dipoles or magnetic-charge magnetic dipoles as long as $[\vE,\vB]$ or $[\vE,\mu_0\vH]$ are chosen as the initial primary fields in free space, respectively.  In fact, for the sake of mathematical simplicity, Maxwell uses ideal continuous differential volume elements of magnetic-charge $\vM$ to define the primary magnetic field $\vH$ and then defines the secondary magnetic field as $\vB=\mu_0(\vH+\vM)$ --- written here in modern SI units \cite[arts. 385, 386]{Maxwell}, \cite{Yaghjian-Reflections}.  {\color{black}For mathematical rigor, it can be assumed that the fields and polarizations are piecewise Lipschitz continuous with possible delta functions in the spatial derivatives across step discontinuities \cite{Liu}.}
\par
Adding the equations that result by crossing $\vE$ or $\vD$ into (\ref{M1a}) and $\vB$ or $\vH$ into (\ref{M1b}) to obtain four possible electromagnetic-field momenta,  then making use of the constitutive relations in (\ref{M2a}), one can obtain an unlimited number of different macroscopic force-momentum density equations depending on the chosen stress dyadic.  Restricting ourselves to five {\color{black}physically reasonable} force-momentum density equations, then integrating them over a volume $V$, one obtains
\begin{subequations}
\lbl{M}
\bea{Ma}
\int\limits_V\left[-(\nabla\cdot\vP)\vE +\left(\nabla\times\vM+\frac{\partial \vP}{\partial t}\right)\times \vB + \eps_0\frac{\partial}{\partial t}(\vE\times\vB)\right]dV\nonumber\\ = \int\limits_V\nabla\cdot\dT_{\!\rm Amp}dV = \int\limits_S \vnh\cdot\dT_{\!\rm Amp} dS
\eea
\bea{Mb}
\int\limits_V\left[\vP\cdot\nabla\vE +\mu_0 \frac{\partial \vP}{\partial t}\times \vH +\mu_0\vM\cdot\nabla\vH -\frac{1}{c^2}\frac{\partial \vM}{\partial t}\times \vE  + \frac{1}{c^2}\frac{\partial}{\partial t}(\vE\times\vH)\right]dV\nonumber\\ = \int\limits_V\nabla\cdot\dT_{\!\rm EL} dV = \int\limits_S \vnh\cdot\dT_{\!\rm EL} dS
\eea
\bea{Mc}
\int\limits_V\left[\vP\cdot\nabla\vE +\mu_0 \frac{\partial \vP}{\partial t}\times \vH +\mu_0\vM\cdot\nabla\vH -\frac{1}{c^2}\frac{\partial \vM}{\partial t}\times \vE -\frac{1}{2}\nabla(\vP\cdot\vE+\mu_0\vM\cdot\vH)  + \frac{1}{c^2}\frac{\partial}{\partial t}(\vE\times\vH)\right]dV\nonumber\hspace{-10mm}\\ = \int\limits_V\nabla\cdot\dT_{\!\rm A} dV = \int\limits_S \vnh\cdot\dT_{\!\rm A} dS\hspace{5mm}
\eea
\bea{Md}
\!\!\int\limits_V\!\left[(\nabla\times\vM)\times\vB +\! \Big(\frac{1}{\eps_0}\nabla\times\vP\Big)\!\times \vD +
\frac{1}{2}\nabla\Big(\frac{1}{\eps_0}\vP\cdot\vD+\vM\cdot\vB\Big)\! - \!\nabla\!\cdot\!\Big(\frac{1}{\eps_0}\vD\vP+\vB\vM\Big) + \frac{\partial}{\partial t}(\vD\times\vB)\right]\!dV\nonumber\hspace{-10mm}\\  = \int\limits_V\nabla\cdot\dT_{\!\rm M}dV = \int\limits_S \vnh\cdot\dT_{\!\rm M} dS \hspace{10mm}
\eea
\bea{Me}
\int\limits_V\left[-\mu_0(\nabla\cdot\vM)\vH +\left(\frac{1}{\eps_0}\nabla\times\vP-\mu_0\frac{\partial \vM}{\partial t}\right)\times \vD+ \mu_0\frac{\partial}{\partial t}(\vD\times\vH)\right]dV\nonumber\\ = \int\limits_V\nabla\cdot\dT_{\!\rm H}dV = \int\limits_S \vnh\cdot\dT_{\!\rm H} dS 
\eea
\end{subequations}
with the respective macroscopic electromagnetic stress dyadics defined as
\begin{subequations}
\lbl{MT}
\be{MTa}
\dT_{\!\rm Amp} = \eps_0\left(\vE\vE -\frac{1}{2}\dI|\vE|^2\right) + \frac{1}{\mu_0}\left(\vB\vB -\frac{1}{2}\dI|\vB|^2\right)
\ee
\be{MTb}
\dT_{\!\rm EL} = \left(\vD\vE -\frac{\eps_0}{2}\dI|\vE|^2\right) + \left(\vB\vH -\frac{\mu_0}{2}\dI|\vH|^2\right)
\ee
\be{MTc}
\dT_{\!\rm A} = \dT_{\!\rm M} = \left(\vD\vE -\frac{1}{2}\dI (\vD\cdot\vE)\right) + \left(\vB\vH -\frac{1}{2}\dI(\vB\cdot\vH)\right)
\ee
\be{MTd}
\dT_{\!\rm H} = \frac{1}{\eps_0}\left(\vD\vD -\frac{1}{2}\dI|\vD|^2\right) + \mu_0\left(\vH\vH -\frac{1}{2}\dI|\vH|^2\right)
\ee
\end{subequations}
where the subscripts $\rm Amp$, $\rm EL$, $\rm A$, $\rm M$, and $\rm H$ stand for the five macroscopic force-momentum equations in (\ref{M}) with the Amperian (as defined in \cite[sec. A.1.6.4]{FCA}), Einstein-Laub \cite[sec. 3]{E&L}, Abraham \cite[eqs. (8) and (Va)]{Abraham1}, \cite[eq. (18b) with the electromagnetic-field momentum density vector defined between eqs. (21) and (22)]{Abraham2}, Minkowski \cite[eqs. (75),(94--97)]{Minkowski}, and Minkowski-with-$\vH$ (as defined herein by the present author) macroscopic electromagnetic-field momentum densities, namely $\eps_0\vE\times\vB$, $\vE\times\vH/c^2$ (for both Einstein-Laub and Abraham), $\vD\times\vB$, and $\mu_0\vD\times\vH$, respectively, and with the corresponding stress dyadics in (\ref{M},\ref{MT}).  Note that the Abraham and Minkowski stress dyadics are identical.  (Four versions of the Poynting theorem can also be expressed with these four electromagnetic-field momenta used as the energy flux vector \cite{Kinsler}.)
\par
The forces on the left-hand sides of (\ref{M}) are given in terms of the polarization densities $\vP$ and $\vM$ and the primary fields as determined by the electromagnetic-field momentum vectors.  The electromagnetic-field momentum densities and electromagnetic stress dyadics labeled as Einstein-Laub, Abraham, and Minkowski are defined by these authors in their original papers.    The designation of ``Amperian'' given to (\ref{Ma}) as well as its electromagnetic-field momentum density and electromagnetic stress dyadic is taken from the textbook by Fano, Chu, and Adler \cite[sec. A.1.6.4]{FCA}.  
I am not aware of any publication by Ampere that contains the equation (\ref{Ma}) or the corresponding electromagnetic-field momentum density and electromagnetic stress dyadic in (\ref{Ma},\ref{MTa}).  Interestingly, Lorentz obtained the macroscopic electromagnetic force in terms of the Einstein-Laub stress dyadic and the Abraham electromagnetic-field momentum before the papers by these authors, but for nonmagnetic material {\color{black} \cite[eqs. (157)--(158)]{LorentzE}}.  The equation in (\ref{Me}) with $\vD\times\vH$ is included here for the sake of completeness in discussing the four alternative forms of the electromagnetic-field momentum densities.
\par
The volume integrals on the left-hand sides of (\ref{M}) \textit{without the electromagnetic-field momentum terms} are the different macroscopic electromagnetic forces ($\vF_{\!\rm Amp}$, $\vF_{\!\rm EL}$,  $\vF_{\!\rm A}$, $\vF_{\!\rm M}$, or $\vF_{\!\rm H}$) corresponding to each formulation.
According to Einstein and Laub \cite{E&L}, the equation containing $\vF_{\!\rm EL}$ and the Abraham macroscopic electromagnetic-field momentum $\vE\times\vH/c^2$ \cite{Abraham1,Abraham2} as well as the stress dyadic $\dT_{\!\rm EL}$ is correct, although Einstein later decided in favor of the Minkowski formulation over the Abraham formulation \cite{EinsteinCP}.  According to Minkowski \cite[eqs. (75),(94--97)]{Minkowski}, the equation with $\dT_{\!\rm M}$ and $\vD\times\vB$ as the macroscopic electromagnetic-field momentum is correct.  \textbf{In fact, any one of these five equations in (\ref{M}) may or may not be correct depending upon whether or not the macroscopic force $\vF_{\!\rm Amp}$, $\vF_{\!\rm EL}$,  $\vF_{\!\rm A}$, $\vF_{\!\rm M}$, or $\vF_{\!\rm H}$ (or some other force) equals the sum of all the electromagnetic microscopic forces in the volume $V$ with its surface $S$ in free space (or in a thin free-space shell surrounding $V$).}
\par
With the surface $S$ in (\ref{M}) lying in a thin free-space shell separating $V$ from the rest of the continuum (see Footnote $1$), \textit{the value of each of the four macroscopic electromagnetic stress dyadics on $S$ is the same, and thus the value of each of their four surface integrals is the same and equals the value of the total \textbf{microscopic} electromagnetic momentum flow across $S$ into $V$, namely $\int_S \vnh\cdot\dcT dS$, because the macroscopic fields equal the microscopic fields in the free-space shell of a sufficiently densely packed dipolar macroscopic continuum}.  This implies that the values of each of the four volume integrals on the left-hand sides of (\ref{M}) are also equal (provided any delta functions in $\nabla\vE$,  $\nabla\vH$, $\nabla\cdot\vP$, $\nabla\cdot\vM$, $\nabla\times\vP$, and $\nabla\times\vM$ at the interface between the material in $V$ and the free space of the thin shell surrounding $V$ are properly included in the volume integrations).\footnote{\lbl{inter}For step functions $u(n)$ in $\vP$ and $\vM$ and delta functions $\delta(n)$ in $\nabla\vE$ and $\nabla\vH$, the integration of the resulting products $u(n)\delta(n)$ at the interface is evaluated using values of $\vP$, $\vM$, $\vE$, and $\vH$ that change rapidly but continuously across the interface in accordance with Maxwell's equations.  This leads to $\int u\delta dn = \int u(du/dn)dn = 1/2$.}  Also, each of the four volume integrals of the divergences of the stress dyadics in (\ref{M}) are equal in value, provided any delta functions in the divergences of the fields of the stress dyadics at the interface between the material in $V$ and the free space of the thin shell surrounding $V$ are properly included in the volume integrations.\footnote{It should be noted that the tangential $\vE$ and $\vH$ fields need not be continuous across the free-space/continuum interface, for example, if the continuum has ``extreme'' constitutive parameters \cite{Yaghjian-Extreme} or if the continuum is strongly spatially dispersive \cite[sec.103]{L&L}, \cite{Silveirinha2009,Yaghjian-sd}.  In strongly spatially dispersive continua, the Poynting vector does not necessarily represent the total energy flow.  Nonetheless, all the force-momentum expressions derived in the present paper hold for temporally and spatially dispersive dipolar continua.  There are no prohibitive restrictions on the constitutive relations except for their satisfying (\ref{M2a}).}  Moreover, because the surface $S$ of $V$ lies in a thin free-space shell, the $\nabla(\vP\cdot\vE+\mu_0\vM\cdot\vH)$ term in (\ref{Mc}) and the $[\nabla(\vP\cdot\vD/\eps_0+\vM\cdot\vB), \nabla\cdot(\vD\vP/\eps_0+\vB\vM)]$ terms in (\ref{Md}) integrate to zero by means of the gradient and divergence integral theorems (since $\vP$ and $\vM$ are zero in free space).  Also, the divergence integral theorem converts the $-(\nabla\cdot\vP)\vE$ and $-\mu_0(\nabla\cdot\vM)\vH$ terms in (\ref{Ma}) and (\ref{Md}) to $\vP\cdot\nabla\vE$ and $\mu_0\vM\cdot\nabla\vH$, respectively. In other words, the five macroscopic force-momentum equations in (\ref{M}) can be reduced to four macroscopic force-momentum equations such that
\bea{4M}
\!\!\!\!\vF_{\!\rm Amp} + \eps_0\frac{d}{dt} \int\limits_V\! \vE\times\vB dV =\vF_{\!\rm EL} + \frac{1}{c^2}\frac{d}{dt} \int\limits_V\! \vE\times\vH dV = \vF_{\!\rm M} + \frac{d}{dt} \int\limits_V\! \vD\times\vB dV = \vF_{\!\rm H} + \mu_0\frac{d}{dt} \int\limits_V\! \vD\times\vH dV\nonumber\hspace{-10mm}\\ = \vF + \eps_0\frac{d}{dt} \int\limits_V \ve\times\vb\, dV = \int\limits_S \vnh\cdot\dT dS = \int\limits_V\nabla\cdot\dT dV\hspace{10mm}
\eea
where, as defined above, $\vF_{\!\rm Amp}$, $\vF_{\!\rm EL}=\vF_{\!\rm A}$, $\vF_{\!\rm M}$, and $\vF_{\!\rm H}$ denote the macroscopic-force volume integrals on the left-hand sides of (\ref{Ma}), (\ref{Mb}), (\ref{Mc}), (\ref{Md}), and (\ref{Me}), respectively, and $\dT$ can be any one of the electromagnetic stress dyadics $\dT_{\!\rm Amp}$, $\dT_{\!\rm EL}$, $\dT_{\!\rm A}=\dT_{\!\rm M}$, $\dT_{\!\rm H}$, or $\dcT$.  The last force-momentum equation in (\ref{4M}) is the microscopic force-momentum equation obtained from (\ref{2}) with the total microscopic force in $V$ given by
\be{4M'}
\vF(t) = \int_V\vf(\vr,t) dV
\ee
where the surface $S$ of $V$ in the microscopic distribution of molecules meanders slightly to avoid cutting through the dipoles so that $S$ encloses a discrete number of dipoles.  By bringing the time derivatives outside of the integrals in (\ref{4M}), it is assumed that $V$ and its surface $S$ do not change with time.  If they do change with time, the partial time derivatives must remain inside the integral signs.
\par
The volume forces in (\ref{M},\ref{4M}) can now be written explicitly as 
\begin{subequations}
\lbl{Mred}
\be{Mreda}
\vF_{\!\rm Amp}(t) = \int\limits_V\left[\vP\cdot\nabla\vE +\left(\nabla\times\vM+\frac{\partial \vP}{\partial t}\right)\times \vB\right] dV
\ee
\be{Mredb}
\vF_{\!\rm EL}(t) = \vF_{\!\rm A}(t) = \int\limits_V\left[\vP\cdot\nabla\vE +\mu_0 \frac{\partial \vP}{\partial t}\times \vH +\mu_0\vM\cdot\nabla\vH -\frac{1}{c^2}\frac{\partial \vM}{\partial t}\times \vE\right] dV
\ee
\be{Mredc}
\vF_{\!\rm M}(t) = \int\limits_V\left[(\nabla\times\vM)\times\vB + \Big(\frac{1}{\eps_0}\nabla\times\vP\Big)\times \vD\right] dV
\ee
\be{Mredd}
\vF_{\!\rm H}(t) =\int\limits_V\left[\mu_0\vM\cdot\nabla\vH +\left(\frac{1}{\eps_0}\nabla\times\vP-\mu_0\frac{\partial \vM}{\partial t}\right)\times \vD\right]dV
\ee
\end{subequations}
where again it is emphasized that any delta functions in the integrands across the free-space/continuum interface of $V$ must be included in the evaluation of the integrals in (\ref{Mred}).
The Einstein-Laub and Abraham electromagnetic forces are equal, which has to be the case, because they have the same electromagnetic-field momentum and with $S$ lying in free space all the electromagnetic-stress dyadic integrals have the same value (as explained above).  
\par
It is now apparent what each of these forces in (\ref{Mred}) represents physically. The Amperian force $\vF_{\!\rm Amp}(t)$ is equal to the sum of the forces exerted by the primary fields $\vE$ and $\vB$ on the electric-charge polarization density $\vP$ (or, alternatively, on the equivalent electric-charge density $-\nabla\cdot\vP$) and on the equivalent electric-current density ($\nabla\times\vM+\partial \vP/\partial t$).  The Einstein-Laub and Abraham forces $\vF_{\!\rm EL}(t)=\vF_{\!\rm A}(t)$ are equal to the sum of the forces exerted by the primary fields $\vE$ and $\vH$ on the electric-charge polarization density $\vP$ (or, alternatively, on the equivalent electric-charge density $-\nabla\cdot\vP$), and the magnetic-charge polarization density $\vM$ (or, alternatively, on the equivalent magnetic-charge density $-\mu_0\nabla\cdot\vM$), and on the electric- and magnetic-polarization current densities $\partial\vP/\partial t$ and $-\mu_0\partial\vM/\partial t$.  The Minkowski force $\vF_{\!\rm M}(t)$ is equal to the sum of the forces exerted by the primary fields $\vD$ and $\vB$ on the equivalent electric- and magnetic-current densities $\nabla\times\vM$ and $\nabla\times\vP/\eps_0$.  The Minkowski-with-$\vH$ force $\vF_{\!\rm H}(t)$ is equal to the sum of the forces exerted by the primary fields $\vD$ and $\vH$ on the magnetic-charge polarization density $\vM$ (or, alternatively, on the equivalent magnetic-charge density $-\mu_0\nabla\cdot\vM$) and on the equivalent magnetic-current density ($\nabla\times\vP/\eps_0-\mu_0\partial\vM/\partial t$).  
\par
For periodic fields, the time derivatives of all four macroscopic electromagnetic-field momenta in (\ref{M},\ref{4M}) average to zero and all the time-averaged macroscopic/microscopic forces are equal to the time-averaged value of the total macroscopic/microscopic electromagnetic momentum flow across $S$ into $V$, that is
\bea{4Mav}
\Big\langle\int\limits_V\nabla\cdot\dT dV\Big\rangle = \int\limits_V\nabla\cdot\langle\dT\rangle dV = \Big\langle\int\limits_S \vnh\cdot\dT dS\Big\rangle = \int\limits_S \vnh\cdot\langle\dT\rangle dS\nonumber\\ = \langle\vF_{\!\rm Amp}\rangle = \langle\vF_{\!\rm EL}\rangle = \langle\vF_{\!\rm A}\rangle  = \langle\vF_{\!\rm M}\rangle = \langle\vF_{\!\rm H}\rangle = \langle\vF\rangle
\eea
where $\langle\,\rangle$ denotes the time average.  Still the surface $S$ of $V$ must lie in free space or in a hypothetical thin free-space shell separating $V$ from the rest of the continuum in order for the time-averaged macroscopic forces in (\ref{4Mav}) to equal the sum of all the time-averaged microscopic forces $\langle\vF\rangle$ on the discrete microscopic dipoles in $V$.
\subsection{Forces on Individual Electric and Magnetic Dipoles}
For general time varying (as opposed to time-averaged periodic) fields, the determination of the total classical electromagnetic force on the microscopic (molecular) electric and magnetic dipole moments $\vp$ and $\vm$ in an electrically small macroscopic volume element $\Delta V$ (with surface $\Delta S$ in free space enclosing $\Delta V$) used to define the macroscopic fields, polarizations, and forces in a dipolar medium requires classical models to represent the molecular electric and magnetic dipole moments. The electrically small $\Delta V$ contain many discrete molecular dipoles and thus it is implicitly assumed that the sources and fields are bandlimited to a maximum frequency $f_{\rm max}$ with a minimum free-space or macroscopic-continuum wavelength $\lambda_{\rm min}$ such that the maximum dimension of $\Delta V$ is much less than $\lambda_{\rm min}$ but with $\Delta V$ still containing many discrete molecular dipole moments.  Thus, the maximum dimension of $\Delta V$ is electrically small ($k_{\rm max}\Delta a\ll 1$, where $k_{\rm max} =2\pi/\lambda_{\rm min}$ and $\Delta a$ is the circumscribing radius of $\Delta V$) and the maximum dimension of the individual molecules is much smaller than the maximum dimension of $\Delta V$.   The electromagnetic fields from thermal motion of the molecules are assumed to produce negligible macroscopic electromagnetic force and momentum within the operational bandwidth.  The molecules can be rotating, translating, and distorting as long as at each instant of time all their multipole moments in each $\Delta V$ are negligible except for electric and magnetic dipole moments, and the accelerations of the molecules are not large enough to produce significant radiation-reaction forces (more precisely, radiation reaction forces of the molecules in $\Delta V$ are of higher order than $\Delta V$, that is, they equal $o(\Delta V)$).  Then at each instant of time, the molecular dipole moments in each $\Delta V$ can be represented by the dipole moments of classical models of electric and magnetic dipoles.  Spatially averaging these dipole moments and their fields at each instant of time using electrically small macroscopic volume elements $\Delta V$ containing large numbers (in principle, approaching an infinite number) of these dipoles yields  macroscopic polarization densities and fields that are well-behaved functions of position and time.
\subsubsection{Electric Dipoles}
Since all electric dipoles of molecules in nature are produced by the separation of electrical charge, all realistic classical electric-dipole models give the same force exerted by external source-free fields $[\vE_e(\vrp,t),\vB_e(\vrp,t)]$ because the total internal force on the electrically small classical electric dipole is zero.  That is, the  quasielectrostatic force exerted on the positive charge by the negative charge is equal and opposite the quasielectrostatic force exerted on the negative charge by the positive charge,  so that the total force exerted on a single electrically small electric dipole $\vp(t)$ equals the force exerted directly by the external fields.  A straightforward derivation of this force from (\ref{fd}) gives \cite[eq. (2.156)]{H&Y}, \cite{Yaghjian-Force}
\be{fed}
\vF_{ed}(t) = \vp(t)\cdot\nabla\vE_e(\vrp,t) + \mu_0\frac{d\vp(t)}{dt}\times \vH_e(\vrp,t)
\ee
with the electrically small electric dipole located at a position $\vrp$ within $\Delta V$ at the time $t$.  The external polarization densities $\vP_e$ and $\vM_e$ are assumed zero at the position of the dipole so that the source-free external electric and magnetic fields at the position of the dipole are related by $\vB_e = \mu_0\vH_e$ and $\vD_e =\eps_0\vE_e$. Note that the second term on the right-hand side of (\ref{fed}) is zero for static electric dipole moments.
\subsubsection{Magnetic Dipoles}
Since all magnetic dipoles of molecules in nature are produced by electric currents (circulating electric charges --- as magnetic charge does not exist),\footnote{Numerous experiments indicate that the intrinsic magnetic dipole moments of elementary particles such as the electron, proton, muon, and neutron are produced by circulating electric currents \cite[p. 191]{Jackson}.} probably the simplest, most appealing, rigorously analyzable stable classical model for the electric-current-produced (Amperian) microscopic magnetic dipole moments of molecules are electrically small perfect electric conductors (PEC's), where the term PEC is used here in the sense of a superconductor whose internal electric and magnetic fields are zero even in the case of static fields \cite{Yaghjian-Force}. Magnetic dipole moments can be induced by externally applied fields on a singly connected PEC (for example, on a PEC sphere) and a stable static magnetic dipole moment can exist on a doubly connected PEC without an externally applied field, for example, on a PEC wire loop.  Notably, Weber and Maxwell \cite[arts. 836--845]{Maxwell} explained both diamagnetism and ordinary magnetism (paramagnetism or ferro(i)magnetism) by means of PEC wire loops with no initial static electric current in the case of diamagnetism, and predominantly initial static electric current in the case of  ordinary magnetism \cite{Yaghjian-Reflections, Yaghjian_PowerEnergy, Yaghjian-Force}.  Of course, electric dipole moments are also induced on PEC's by externally applied fields.
\par
It is rigorously proven from (\ref{fd}) in \cite{Yaghjian-Force} that, remarkably, when \textit{arbitrarily time varying} external fields are applied to electrically small PEC's (for example, PEC wire loops), \textit{no matter how electrically small the PEC's}, there is, in addition to a direct external electromagnetic-field force, an internal ``hidden momentum" electromagnetic force, namely $(-1/c^2)\partial [\vm(t)\times\vE_e(\vrp,t)]/\partial t$, induced indirectly by the external fields, where $-\vm(t)\times\vE_e(\vrp,t)$ is the microscopic hidden momentum of the Amperian magnetic dipole moment.  \textbf{The proof in \cite{Yaghjian-Force} is crucial because it is the only rigorous derivation of $-\vm\times\vE$ as the hidden momentum for Amperian dipoles subject to arbitrarily time varying external fields.}\footnote{Unfortunately, the term ``hidden momentum'' has a somewhat mysterious connotation, whereas this momentum simply arises from the force exerted on the electric charge-current of the PEC by the internal fields produced by the same electric charge-current \cite{Yaghjian-Force}.   Some authors, such as Boyer \cite{Boyer1,Boyer2}, prefer the term ``internal momentum'' to ``hidden momentum.''  Although not mysterious, it may be nonetheless surprising that this self-force internal momentum does not vanish for electrically small PEC's.  The reason for this nonvanishing self force is that the electromagnetic quasistatic fields of the PEC do not uncouple into quasistatic electric fields and quasistatic magnetic fields.  The electromagnetic fields remain coupled to produce a ``hidden momentum'' force even as the size of the PEC approaches zero \cite{Yaghjian-Force}.   Also, the derivation in \cite{Yaghjian-Force} of (\ref{fmd})  applies to electrically small conductors with finite conductivity as well as PEC's, except that conductors with finite conductivity cannot support an intrinsic static magnetic dipole moment.}  Thus, the total force exerted by the external fields on a single electrically small PEC Amperian magnetic dipole moment $\vm(t)$ located at the position $\vrp$ is given by \cite[eq. (2.163)]{H&Y}, \cite{Yaghjian-Force}
\be{fmd}
\vF_{md}(t) = \mu_0\vm(t)\cdot\nabla\vH_e(\vrp,t) -  \frac{1}{c^2} \frac{d\vm(t)}{dt}\times \vE_e(\vrp,t).
\ee
This is the same force that would be exerted directly by the external fields on an electrically small magnetic-charge magnetic dipole moment $\vm(t)$ \cite[eq. (2.166)]{H&Y} (if magnetic charge existed), for which the internal forces cancel like those of the electric-charge electric dipole.  Indeed, it is the same force that Einstein and Laub \cite{E&L} found assuming magnetic-charge models for the magnetic dipoles.   (They didn't consider Amperian magnetic dipoles in their paper.)  The ``hidden-momentum" electromagnetic force on an Amperian magnetic dipole has resulted in its total force exerted by the external electric and magnetic fields equal to the force that would be experienced by a magnetic-charge magnetic dipole \cite{Yaghjian-Force}.  Note that the second term on the right-hand side of (\ref{fmd}) is zero for static magnetic dipole moments.
\par
The molecules of the dipolar material with different dipole moments $\vp(t)$ and $\vm(t)$ located at different positions $\vrp$ at time $t$ can be moving within the aforementioned acceleration limits that prevent significant radiation-reaction forces (as long as $\vp(t)$ and $\vm(t)$ are the instantaneous dipole moments in the laboratory frame of reference).   Even if all the molecular dipoles are modeled by lossless scatterers, when brought together to form a macroscopic continuum, energy losses can be exhibited in the continuum by, for example, assuming the discrete lossless dipoles are connected to one another and to the other molecules of the continuum by lossy, linear or nonlinear, compressible and torsional springs.  {\color{black}Thermal losses can occur within the springs.}  However, as mentioned above, the macroscopic electromagnetic forces and momenta of thermal motion are assumed negligible within the operational bandwidth.  Although the molecules can be moving within the aforementioned acceleration limits that prevent significant radiation-reaction forces, at any one time $t$, each of the defining electrically small volume elements $\Delta V$ is chosen to contain a discrete number of dipoles (so that each $\Delta S$ lies in free space).  Of course, if the dipolar continuum is moving with constant velocity in free space, the dipole moments, forces, and momenta may be determined more conveniently in an inertial reference frame moving with the dipolar continuum, and the corresponding forces and momenta in the laboratory frame can be found from the Lorentz relativistic transformations.
\par
The sum of the electric and magnetic dipole forces in (\ref{fed}) and (\ref{fmd}) is 
\be{fedpm}
\vF_{d}(t)\! = \!\vF_{ed}(t)+\vF_{md}(t)\! =\! \vp(t)\cdot\nabla\vE_e(\vrp,t) + \mu_0\frac{d\vp(t)}{dt}\!\times \vH_e(\vrp,t)
+ \mu_0\vm(t)\cdot\nabla\vH_e(\vrp,t) -  \frac{1}{c^2} \frac{d\vm(t)}{dt}\!\times \vE_e(\vrp,t).
\ee
The main limitations of the expression for the force in (\ref{fedpm}) are that it neglects the quadrupole and higher-order multipole moments of the charge-current distribution that produces the dipole moments $\vp$ and $\vm$; and, moreover, the magnetic dipole moment $\vm$ depends on the origin of the coordinate system (unless $\vp=0$) such that a displacement $\Delta \vr$ of the origin changes the magnetic dipole moment by $(\Delta\vr\times d\vp/dt)/2$ \cite[p. 10]{H&Y}.  Both these limitations imply that the fractional error in the magnitude of the force in (\ref{fedpm}) is $O(k_{\rm max} a)$ where $a$ is the radius of the sphere that circumscribes the charge-current sources of $\vp$ and $\vm$ \cite[pp. 3,8,11,13]{R&D}.
\par
For time-harmonic ($e^{-i\w t}$) dipoles and fields, the time-average of the force in (\ref{fedpm}) can be written as
\be{fta}
\langle\vF_{d}\rangle = \frac{1}{2}{\rm Re}\left[\vp_\w\cdot\nabla\vE_{e\w}^* -i\w\mu_0\vp_\w \times \vH_{e\w}^*
+ \mu_0\vm_\w\cdot\nabla\vH_{e\w}^* + i\w \mu_0\eps_0\vm_\w\times \vE_{e\w}^*\right]
\ee
where the subscript $\w$ denotes the frequency of the time-harmonic vectors.  With the dipoles lying outside the sources of the external fields, vector identities and the Maxwell curl equations combine to reduce (\ref{fta}) to
\be{ftar}
\langle\vF_{d}\rangle = \frac{1}{2}{\rm Re}\left[(\nabla\vE_{e\w}^*)\cdot\vp_\w 
+ \mu_0(\nabla\vH_{e\w}^*)\cdot\vm_\w \right].
\ee
This expression for the time-averaged harmonic force on dipoles has been obtained in \cite{Nieto,Chaumet} by evaluating the stress dyadic of the fields of electrically small ($ka \ll 1,\; k=\w/c$ and $a$ is the radius of the circumscribing sphere of the particle) magnetodielectric particles illuminated by external fields.  The expressions in \cite{Nieto,Chaumet} also include a term equal to $k^4\sqrt{\mu_0\eps_0}\,\vp\times\vm^*/(6\pi)$, which, however, is $O[(ka)^3]$ times the dipolar forces in (\ref{ftar}) for magnetodielectric dipoles induced by external fields and thus would generally be negligible compared to forces arising from the origin-dependence of $\vm$ and the quadrupole moments induced by the external fields in the magnetodielectric particles.  As mentioned above, these neglected forces are $O(ka)$ times the dipolar forces in (\ref{ftar}), and they were not included in the derivation in \cite{Nieto,Chaumet}, which assumed origin-independent $\vm$ and only dipole fields in the stress dyadic.  Consequently, the $k^4\sqrt{\mu_0\eps_0}\,\vp\times\vm^*/(6\pi)$ term is superfluous.
\subsection{Macroscopic Dipolar Forces Determined from Microscopic Dipole Forces}
To obtain an expression for the macroscopic dipolar forces from the microscopic electric-dipole and magnetic-dipole forces in (\ref{fed}) and (\ref{fmd}), consider a distribution of many of these dipoles at each instant of time $t$ in a defining electrically small macroscopic volume element $\Delta V$ with its surface $\Delta S$ in free space surrounding $\Delta V$; see Footnote $1$.  {\color{black}The forces between all the isolated electrically small electric-charge electric dipoles and the PEC combined electric and magnetic dipoles in $\Delta V$ are quasistatic forces that cancel.}\footnote{{\color{black}The electric and magnetic dipole moments $\vp$ and $\vm$ induced by external fields on PEC's are not isolated, but the force on the electrically small PEC has already been proven \cite{Yaghjian-Force}, as explained above, to be equal to the sum of the electric- and magnetic-dipole forces in (\ref{fed}) and (\ref{fmd}).}  An alternative heuristic way of showing that the isolated molecules in $\Delta V$ do not exert a net force on each other is to consider each molecule lying in a small spherical hole within a spherical $\Delta V$ of uniform polarization which is separated by a thin free-space spherical shell from the rest of the macroscopic continuum.  For an electrically small $\Delta V$, the quasistatic fields in the hole produced by the polarization in $\Delta V$ will be zero (the uniform fields within the hole that are produced by the surface polarization charges and currents on the surfaces of $\Delta V$ and the hole cancel to $o(\Delta V)$ \cite{Yaghjian-Green}) and thus no force will be exerted by the polarization in $\Delta V$ on the dipole moments of the molecules within the hole.}   In addition, the radiation-reaction forces of the molecules in $\Delta V$  are assumed to be of higher order than $\Delta V$, that is, $o(\Delta V)$.   Thus only the external fields from sources outside $\Delta V$ will produce an appreciable net electromagnetic force on the dipoles in $\Delta V$.  The spatially averaged electric polarization of all the $\vp$'s in $\Delta V$ is $\vP(\vrp,t)$ so that the external force on this electric polarization (all the discrete dipole moments $\vp$) in $\Delta V$ is given by (\ref{fed}) with $\vP\Delta V$ substituted for $\vp$, namely
\be{fedDV}
\Delta\vF_{ed}(\vrp,t) = \left[\vP(\vrp,t)\cdot\nabla\vE_e(\vrp,t) + \mu_0\frac{\partial\vP(\vrp,t)}{\partial t}\times \vH_e(\vrp,t)\right]\Delta V
\ee
where $\vrp$ is a point in $\Delta V$.  
\par
The external electric and magnetic fields in (\ref{fedDV}) are the fields in $\Delta V$ with $\vP$ and $\vM$ in $\Delta V$ removed, that is, the cavity fields, $\vE_e=\vE_c = \vE - \vE_s$ and $\vH_e=\vH_c = \vH - \vH_s$ where $\vE$ and $\vH$ are the spatially averaged macroscopic electric and magnetic fields and $\vE_s$ and $\vH_s$ are the electric and magnetic fields produced by the equivalent surface charge and current densities $\vnh\cdot\vP$ and $\vM\times\vnh$ at the outer surface of the free-space shell that surrounds the electrically small $\Delta V$ \cite[p. 46]{H&Y}.  (Incidentally, for spherical $\Delta V$, the $\vE_s$ and $\vH_s$ are approximately uniform throughout $\Delta V$ so that $\nabla\vE_s$ and $\nabla\vH_s$ are approximately zero throughout $\Delta V$.)  Because these surface charge and current densities have equal and opposite counterpart surface charge and current densities for the volume elements adjacent to $\Delta V$, they create equal and opposite forces on the electrically small $\Delta V$ and its adjacent volume elements that cancel when the forces on all the $\Delta V$ in a volume $V$ are summed to obtain the integrated force in the dipolar material in $V$\footnote{The volume elements $\Delta V$ that border the polarization/free-space interface of $V$ have one side without an adjacent volume element.  This merely changes the values of $\vE_s$ and $\vH_s$ but not the cancellation argument leading from (\ref{fedDV}) to (\ref{fedDV'}).  However, it is emphasized that the {\color{black}uncanceled} equivalent surface charge/current densities at the surface of $V$ contribute to the force in (\ref{Fela}) on the volume $V$ of dipolar material since these forces are taken into account by delta functions in $\nabla\vE$ and $\nabla\vH$ in the integrand of (\ref{Fela}).}; thus the $\vE_s$ and $\vH_s$ fields can be ignored and (\ref{fedDV}) becomes effectively for volume elements inside the dipolar macroscopic continuum
\be{fedDV'}
\Delta\vF_{ed}(\vrp,t) = \left[\vP(\vrp,t)\cdot\nabla\vE(\vrp,t) + \mu_0\frac{\partial\vP(\vrp,t)}{\partial t}\times \vH(\vrp,t)\right]\Delta V
\ee
a relatively uncomplicated well-defined expression for the macroscopic force density, which when integrated to get the force on the electric polarization density $\vP$ in a volume $V$, gives the sum of the microscopic forces on all the discrete dipole moments $\vp$ in $V$.
\par
Similarly, (\ref{fmd}) yields the macroscopic force on the macroscopic magnetic polarization (magnetization) $\vM$ in $\Delta V$
\be{fmdDV}
\Delta\vF_{md}(\vrp,t) = \left[\mu_0\vM(\vrp,t)\cdot\nabla\vH_e(\vrp,t)  - \frac{1}{c^2} \frac{\partial\vM(\vrp,t)}{\partial t}\times \vE_e(\vrp,t)\right]\Delta V
\ee
that equals the sum of the microscopic forces on the discrete magnetic dipole moments $\vm$ in $\Delta V$.
\par
By the same argument that led from (\ref{fedDV}) to (\ref{fedDV'}), the force density in (\ref{fmdDV}) can be re-expressed effectively as
\be{fmdDV'}
\Delta\vF_{md}(\vrp,t) = \left[\mu_0\vM(\vrp,t)\cdot\nabla\vH(\vrp,t)  - \frac{1}{c^2} \frac{\partial\vM(\vrp,t)}{\partial t}\times \vE(\vrp,t)\right]\Delta V
\ee
for the purpose of integrating over a volume $V$ to get the macroscopic force on $\vM$ equal to the sum of the microscopic forces on the discrete magnetic dipole moments $\vm$ in $V$.  It is emphasized that (\ref{fmdDV})--(\ref{fmdDV'}) holds for both diamagnetic and paramagnetic/ferro(i)magnetic magnetization $\vM$.
\par
Within a volume $V > \Delta V$ with the surface $S$ of $V$ lying in a thin free-space shell, the volume elements in (\ref{fedDV'}) and (\ref{fmdDV'}) used to define the electromagnetic forces in $V$ can change shape slightly and be shifted so that $\vr$ can take any value in $V$.  (Because each $\Delta V$ contains a discrete number of dipoles, there will be a small spatial jitter in this macroscopic force density with the jitter becoming smaller with the larger the number of dipoles per unit volume.  This jitter can be smoothed by various techniques such as moving averages.)  Thus the total macroscopic electromagnetic force $\vF_{\!e\ell}(t)$ on the macroscopic polarization densities $\vP$ and $\vM$ in a volume $V$ is determined approximately by integrating the sum of the macroscopic electric-dipole and magnetic-dipole force densities in (\ref{fedDV'}) and 
(\ref{fmdDV'}) to get
\begin{subequations}
\lbl{Fel}
\be{Fela}
\vF_{\!e\ell}(t)\! = \!\int\limits_V \left[\vP(\vrp,t)\cdot\nabla\vE(\vrp,t) + \mu_0\frac{\partial\vP(\vrp,t)}{\partial t}\times \vH(\vrp,t) + \mu_0\vM(\vrp,t)\cdot\nabla\vH(\vrp,t)  - \frac{1}{c^2} \frac{\partial\vM(\vrp,t)}{\partial t}\times \vE(\vrp,t)\right]dV
\ee
which, with the surface $S$ of $V$ in free space and any delta functions in the spatial derivatives across the free-space/continuum interface of $V$ properly included in the integrations (see Footnote $2$), equals the sum of the microscopic forces on the discrete dipole moments $\vp$ and $\vm$ in $V$, that is
\be{Fela'}
\vF_{\!e\ell}(t) = \vF(t).
\ee
(The integration in (\ref{Fela}) becomes a more accurate representation of the sums in (\ref{fedDV'}) and (\ref{fmdDV'}) as the number of dipoles per cubic minimum wavelength becomes larger and $\Delta V$ becomes electrically smaller.)   We see from (\ref{Mredb}) and (\ref{Fela},\ref{Fela'}) that
\be{Felb}
\vF_{\!e\ell}(t) = \vF_{\!\rm EL}(t) = \vF(t).
\ee
\end{subequations}
Since we have proven that the macroscopic electromagnetic force $\vF_{\!e\ell}=\vF_{\!\rm EL}$ equals the sum of the microscopic electromagnetic forces in $V$, namely $\vF$ in (\ref{4M'}), we also have from (\ref{4M}) that
\be{eqeb}
\frac{1}{c^2}\frac{d}{dt}\int\limits_V \vE\times\vH\, dV = \eps_0\frac{d}{dt}\int\limits_V \ve\times\vb \,dV
\ee
that is, the macroscopic and microscopic electromagnetic-field momenta in $V$ are equal.  If the total electromagnetic momentum entering the volume is zero, that is,  
$\int_S \vnh\cdot\dT dS = 0$ (for example, if the surface $S$ lies outside all the fields), then (\ref{4M}) and (\ref{Felb}) show that the macroscopic force on the volume $V$ of material is given by
\be{Felz}
\vF_{\!e\ell}(t) = \vF_{\!\rm EL}(t) = \vF(t) = -\frac{1}{c^2}\frac{d}{dt}\int\limits_V \vE\times\vH\,dV \neq -\frac{d}{dt}\int\limits_V \vD\times\vB\,dV.
\ee
\textit{That is, we have shown that, with no net electromagnetic momentum crossing $S$, the macroscopic electromagnetic force exerted by an electromagnetic pulse on a volume $V$ of dipolar material is given by the negative time derivative of the Abraham macroscopic electromagnetic-field momentum and not by the negative time derivative of the Minkowski (or any other) macroscopic electromagnetic-field momentum.}  Integrating the equality in (\ref{Felz}) over a time interval shows that the macroscopic electromagnetic impulse applied to the charge-current in this volume $V$ plus the change in the Abraham macroscopic electromagnetic-field momentum in $V$ during that time interval is zero if no net electromagnetic momentum enters or leaves the volume $V$.
\par
The macroscopic electromagnetic force in (\ref{Fel}) is the Einstein-Laub \cite{E&L} macroscopic electromagnetic force in (\ref{Mb},\ref{Mredb}) with the associated Abraham/Einstein-Laub electromagnetic-field momentum $\vE\times\vH/c^2$ in (\ref{Mb},\ref{Mc})).  However, Einstein and Laub simply assume microscopic  magnetic-charge magnetic dipoles analogously to microscopic electric-charge electric dipoles even though magnetic dipoles in matter are produced by circulating electric currents.  They then generalize without proof their force expressions on microscopic electric and magnetic dipoles to the corresponding expressions with macroscopic electric and magnetic polarization densities and fields.  Also, Einstein and Laub did not indicate or speculate on the possible existence of internal momentum (today referred to as hidden momentum) induced by the external fields applied to Amperian (circulating-electric-current) magnetic dipoles.  In fact,  Einstein \cite[p. 591]{EinsteinCP} later wrote that the electromagnetic tensor that he and Laub had derived was ``wrong" and that the correct tensor was that of Minkowski.  
\par
Mansuripur \cite{Mansuripur2017} also simply assumes that ``magnetism is no longer associated with an electric current density, but rather with bound magnetic-charge and bound magnetic-current densities" to argue for the Einstein-Laub macroscopic force (and the associated Abraham macroscopic electromagnetic-field momentum) but, again, without proving that the macroscopic force is equal to the sum of the microscopic electromagnetic forces on realistic models of electric and magnetic dipoles in $V$.  On the one hand, assuming magnetism is produced by magnetic-charge separation denies the experimental results of modern physics that magnetic charge and magnetic-charge dipoles do not exist in nature; and on the other hand,  avoids treating the problem of magnetic polarization properly because of the past difficulties with dealing rigorously with realistic models of Amperian magnetic dipoles.   
\par
Here in the present work, we have rigorously proven that the macroscopic electromagnetic force in (\ref{Mredb},\ref{Fel}) on the volume $V$ of macroscopic electric and magnetic polarization $\vP$ and $\vM$ is equal to the sum of the electromagnetic forces on the discrete microscopic electric-charge electric dipoles and electric-current (Amperian) magnetic dipoles in $V$ using realistic classical models for the discrete microscopic electric and magnetic dipoles.  The macroscopic force in (\ref{Mredb},\ref{Fel}) was first obtained in \cite[eq. 2.173]{H&Y} using the particular example of a PEC sphere to argue for the hidden momentum from arbitrary time varying fields rather than the general proof given in \cite{Yaghjian-Force} for the hidden-momentum force on arbitrarily shaped microscopic PEC models of magnetic dipoles.  \textit{\textbf{In summary, it is proven that for these realistic classical models of electric and magnetic dipoles found in nature, the physically meaningful macroscopic electromagnetic force and momentum  ({\color{black}that is, those} equal to the sum of the microscopic-dipole electromagnetic forces and momenta, {\color{black}respectively}) are the Einstein-Laub macroscopic electromagnetic force $\vF_{\!\rm EL}(t)$ in (\ref{Mredb},\ref{Fel}) and the Abraham/Einstein-Laub macroscopic electromagnetic-field momentum density $\vE\times\vH/c^2$}}.  {\color{black}They are related to the integrals of the stress dyadics as given in (\ref{4M}), namely
\be{4Mone}
\vF_{\!\rm EL}(t) + \frac{1}{c^2}\frac{d}{dt} \int\limits_V\! \vE\times\vH dV = \int\limits_S \vnh\cdot\dT dS = \int\limits_V\nabla\cdot\dT dV.
\ee}\noindent
Also, as explained above, the macroscopic polarization densities $\vP$ and $\vM$ in (\ref{Fela}) can be lossless or lossy, independent of whether the microscopic models of the discrete dipoles are lossless or lossy, and their constitutive relations are unrestricted except for obeying (\ref{M2a}).
\par
The rigorously derived macroscopic force density in (\ref{Fela}) shows that internal (hidden) momentum force is produced by the applied fields in macroscopic magnetization $\vM$ as well as in the microscopic momentum of the Amperian dipoles comprising the magnetization, and, in particular, in  the magnetization of artificial molecules (inclusions) of metamaterials containing magnetic (or magnetodielectric) material.  That is, the force density in the magnetization $\vM$ of inclusions is given by $-\partial\vM /\partial t \times \vE/c^2$ rather than $\vM\times(\partial \vE/\partial t)/c^2$ and, thus,  includes a macroscopic hidden momentum $-\partial(\vM\times\vE)/(c^2\partial t)$.  This macroscopic hidden momentum for the magnetization $\vM$ is a consequence of the microscopic hidden momentum which is rigorously found {\color{black}\cite{Yaghjian-Force}} for the classical conductor Amperian models of the discrete magnetic dipoles comprising $\vM$ and is not lost in the averaging process required to obtain the macroscopic magnetization $\vM$ from the microscopic Amperian magnetic dipole moments $\vm$.
\section{Kinetic and Canonical Momenta}\lbl{Canon}
It is emphasized that the force $\vF_{\!e\ell}(t)= \vF_{\!\rm EL}(t)$ in (\ref{4M},\ref{Mredb},{\ref{Fel}) equals the total electromagnetic force on realistic models of the microscopic dipoles in $V$.  If the dipoles are rigidly attached to a rigid lattice structure or any other rigid material with a fixed volume inside $V$ whose surface $S$ in free space surrounds all of the material, then this electromagnetic force is transferred to the rigid material.  Also, if the rigid material inside $V$ is held fixed, the opposite force must be exerted by whatever outside agent (other than the given applied electromagnetic fields) is holding the rigid material inside $V$ fixed.  On the other hand, even if the outer surface of the material inside $V$ is held fixed by an outside agent, but the charge carriers accelerate appreciably (yet not enough to produce radiation-reaction forces greater than $o(\Delta V)$) with respect to the fixed  surface of material inside $V$ or they collide with other molecules of the material that can accelerate with respect to its fixed surface, producing a total change in kinetic momentum $\vG_{\rm k}(t)$ of the material inside $V$, then the force $\vF_{\!ag}(t)$ exerted by the outside agent holding the surface of the material inside $V$ fixed will satisfy Newton's relativistic equation of motion 
\be{agent1}
\vF_{\!ag}(t) + \vF_{\!e\ell}(t) = \vF_{\!ag}(t) + \vF_{\!EL}(t) = \frac{d\vG_{\rm k}(t)}{dt}
\ee
{\color{black}provided the total radiation reaction force of the accelerating macroscopic polarization is negligible \cite{Yaghjian-Springer}.}
 Thus $\vF_{\!ag}(t)$ can be expressed with the help of {\color{black}(\ref{4Mone})} as
\be{agent2}
\vF_{\!ag}(t) = \frac{d\vG_{\rm k}(t)}{dt} +\frac{1}{c^2} \frac{d}{dt}\int\limits_V \vE\times\vH \,dV   - \int\limits_S \vnh\cdot\dT dS 
\ee
where $d\vG_{\rm k}(t)/dt$ is the time rate of change of the kinetic momentum of all the material inside the volume $V$.  If outside the surface $S$ of $V$ there is only free space, the volume $V$ in (\ref{agent2}) can be replaced by the volume $V_\infty$ of all space such that $\dT$ is zero on $S_\infty$ for a finite pulse.  Then (\ref{agent2}) becomes simply
\be{agent3}
\vF_{\!ag}(t) = \frac{d\vG_{\rm k}(t)}{dt} +\frac{1}{c^2} \frac{d}{dt}\int\limits_{V_\infty} \vE\times\vH \,dV.    
\ee
Silveirinha \cite[eq. (10b)]{Silveirinha} obtains a form of (\ref{agent2}) with the microscopic electromagnetic-field momentum replacing the macroscopic electromagnetic-field momentum in (\ref{agent2}) but without proving which macroscopic force (namely, $\vF_{\!\rm EL}$) equals the sum of the microscopic forces, or which macroscopic electromagnetic-field momentum (namely, $\vE\times\vH/c^2$) equals the sum of the microscopic electromagnetic-field momenta.  For the Minkowski formulation (which gives neither the correct macroscopic electromagnetic force nor the correct electromagnetic-field momentum), $\vE\times\vH/c^2$ in the volume integral of (\ref{agent2}) would be replaced by $\vD\times\vB$.
\par
In general, $\vF_{\!ag}(t)$ in (\ref{agent1}--\ref{agent3}) is the total outside-agent force exerted on the material inside any volume $V$ outside of which exists only free space even if the material inside $V$ is allowed to accelerate and deform, as long as the macroscopic fields and polarizations are those in the moving material inside $V$ (as seen in the laboratory frame) and $\vG_{\rm k}(t)$ is the total change in kinetic momentum of the material inside $V$.  The momentum $\vG_{\rm k}(t)$ can be considered as a macroscopic kinetic momentum equal to the change in the total kinetic momentum of all the material (charged and uncharged particles) inside $V$.  This change in total kinetic momentum is brought about by the applied electromagnetic and outside-agent forces.   The equation (\ref{agent2}) says that the outside-agent force exerted on the material inside the volume $V$  equals the time rate of change of the kinetic plus electromagnetic-field momentum inside $V$ plus the time rate of change of the total electromagnetic momentum entering $V$; recall that $\vnh$ is the unit normal pointing out of $V$.  Similarly, (\ref{agent3}) says that the outside-agent force exerted on the material in the volume $V$ equals the time rate of change of the kinetic momentum of the material in $V$ plus the time rate of change of the electromagnetic-field momentum throughout all space $V_\infty$.   Unfortunately, $d\vG_{\rm k}(t)/dt$ is usually unknown and may be difficult to determine, although in some cases where the material has a high rigidity and an outside-agent force keeps the surface of the volume $V$ of material fixed, it may be reasonable to assume $d\vG_{\rm k}(t)/dt$ is negligible.  If no outside force is supplied by an agent to the material inside the volume $V$, then $\vF_{\!ag}(t)=0$  and we have from (\ref{agent1}--\ref{agent3})
\be{agent0}
\vF_{\!\rm EL}(t) = \frac{d\vG_{\rm k}(t)}{dt} = \int\limits_S \vnh\cdot\dT dS -\frac{1}{c^2} \frac{d}{dt}\int\limits_V \vE\times\vH dV = -\frac{1}{c^2} \frac{d}{dt}\int\limits_{V_\infty} \vE\times\vH \,dV.    
\ee
\par
Although it has been shown that the Einstein-Laub macroscopic force and Abraham/Einstein-Laub macroscopic electromagnetic-field momentum are the ones that equal the sum of the microscopic forces and the sum of the microscopic electromagnetic-field momenta, {\color{black}respectively,} it may be revealing to relate the Minkowski and Einstein-Laub macroscopic forces and electromagnetic-field momenta.  Toward this end, use (\ref{4M}) to write
\be{km1}
\vF_{\!\rm EL}(t)-\vF_{\!\rm M}(t) = \frac{d}{dt} \int\limits_V \vD\times\vB\, dV  
-\frac{1}{c^2} \frac{d}{dt}\int\limits_V \vE\times\vH dV.
\ee
If the outside-agent force is zero, $\vF_{\!\rm EL}=d\vG_{\rm k}/dt$ and (\ref{km1}) can be rewritten as 
\be{km2}
\frac{d\vG_{\rm k}(t)}{dt}-\vF_{\!\rm M}(t) = \frac{d}{dt}\int\limits_V \vD\times\vB\, dV  
-\frac{1}{c^2}\frac{d}{dt}\int\limits_V \vE\times\vH dV.
\ee
Furthermore, expressing the Minkowski force in terms of the time rate of change of a ``Minkowski-force momentum,'' that is
\be{km3}
\vF_{\!\rm M}(t) = \frac{d\vG_{\rm M}(t)}{dt}
\ee
one obtains
\be{km4}
\frac{d\vG_{\rm k}(t)}{dt}-\frac{d\vG_{\rm M}(t)}{dt}  = 
  \frac{d}{dt} \int\limits_V \vD\times\vB\, dV - \frac{1}{c^2} \frac{d}{dt}\int\limits_V \vE\times\vH \,dV
\ee
or, assuming the initial macroscopic fields are zero and $\vG_{\rm k}(t)$ is the change in kinetic momentum from the initial kinetic momentum, and $\vG_{\rm M}(0)=0$, then
\be{km5}
\vG_{\rm k}(t)-\vG_{\rm can}(t) =  \int\limits_V \left(\vD\times\vB -  \vE\times\vH/c^2\right) dV =  \int\limits_V \left(\vP\times\vB - \vM\times\vE/c^2\right) dV
\ee
where the Minkowski-force momentum $\vG_{\rm M}(t)$ has been renamed $\vG_{\rm can}(t)$ to 
correspond to the microscopic ``canonical momentum'' of Lembessis et al. \cite{Lembessis,Baxter} and to the macroscopic ``medium canonical momentum'' of Barnett and Loudon \cite{Barnett, B&L}.  Specifically, (\ref{km5}) corresponds to the macroscopic equation \cite[eq. 4.4]{B&L} and to the microscopic equation  \cite[eq. (50)]{Lembessis}, where it should be noted that \cite{Lembessis} treats only single discrete electric dipoles (and no magnetic dipoles).
\par
In the context of the macroscopic formulation of the present paper, the vector $(\vP\times\vB -  \vM\times\vE/c^2)$ in (\ref{km5}) is the sum of the macroscopic hidden momentum density ($-\vM\times\vE/c^2$) of Amperian (circulating-electric-current) magnetic dipoles and the macroscopic hidden momentum density ($\vP\times\vB$) of hypothetical circulating-magnetic-current electric dipoles replacing the electric-charge electric dipoles.\footnote{The $\vP\times\vB$ can be shown to be the macroscopic hidden momentum density of hypothetical circulating-magnetic-current electric dipoles in the same way that it is shown from \cite{Yaghjian-Force} that $-\vM\times\vE/c^2$ is the macroscopic hidden momentum density of  circulating-electric-current magnetic dipoles.  In the absence of any outside-agent force, the kinetic momentum is simply equal to the electromagnetic-force momentum imparted to the dipoles by the applied fields [$\vF_{\!EL}(t)=\vF_{\!e\ell}(t) = d\vG_{\rm k}(t)/dt$].  Therefore, the canonical (Minkowski-force) momentum is simply the direct momentum (that is, the electromagnetic momentum without the internal ``hidden'' momentum) imparted by the applied fields to electric and magnetic dipole moments created by magnetic and electric circulating current, respectively [$\vG_{\rm M}(t)=\vG_{\rm can}(t) = \vG_{\rm k}(t)- \int_V (\vP\times\vB - \vM\times\vE/c^2)dV$], where $\int_V (\vP\times\vB - \vM\times\vE/c^2)dV$ is the internal ``hidden'' momentum of the magnetic- and electric-circulating-current dipoles in $V$.}  Thus, the difference between the Abraham and Minkowski electromagnetic forces in (\ref{km1}) equals the sum of the hidden-momentum forces on these electric-current magnetic dipoles and hypothetical magnetic-current electric dipoles.  This makes sense because the $\vE$ and $\vH$ primary fields of the Abraham formulation require electric- and magnetic-charge electric and magnetic dipoles, respectively, (that exhibit no hidden momentum) to derive the Maxwell equations and constitutive relations in (\ref{M1},\ref{M2a}), whereas the $\vD$ and $\vB$ primary fields of the Minkowski formulation require electric- and magnetic-current magnetic and electric dipoles, respectively, (that exhibit the hidden momentum in (\ref{km5})) to derive the Maxwell equations and constitutive relations in (\ref{M1},\ref{M2a}). However, neither (\ref{km5}) nor the references \cite{Lembessis,Baxter,Barnett,B&L} prove, as is done in the above analysis of the present paper, that it is the time derivative of the Abraham/Einstein-Laub macroscopic electromagnetic-field momentum, and not the time derivative of the Minkowski macroscopic electromagnetic-field momentum, along with any one of the stress dyadics in (\ref{3},\ref{MT}), that determine the correct macroscopic electromagnetic force $\vF_{\rm EL}$ on a volume of bulk dipolar material (namely the force that equals the sum of the microscopic electromagnetic forces on realistic discrete electric-charge electric dipoles and electric-current (Amperian) magnetic dipoles in that volume of material).  It should also be emphasized that in the context of classical physics, the canonical momentum is defined by (\ref{km1}--\ref{km5}) and \cite[eq. 4.4]{B&L}, whereas fundamentally the ``canonical momentum'' is a concept that arises in quantum electrodynamical scattering of light by an atom \cite{Lembessis,Baxter}.  Lembessis et al. \cite{Lembessis,Baxter} refer to the microscopic quantum electrodynamical canonical momentum as a ``R\"ontgen-type interaction term.''  However, none of these papers, or any other previous papers, as far as I am aware, have associated the difference between the kinetic and canonical momenta with the sum of the internal ``hidden'' electromagnetic momentum of electric-current magnetic dipoles and magnetic-current electric dipoles.
\section{Experimental Verification of the Einstein-Laub/Abraham Formulation}
Through the years there have been many experiments done to measure the time-averaged macroscopic electromagnetic radiation force, especially the force on mirrors and on the surfaces or membranes between two fluids.  None of the time-averaged force experiments can distinguish between the Einstein-Laub/Abraham formulation and the Minkowski formulation because there is no difference between the two for time-averaged forces, as shown in Section \ref{Macro}.
\par
However, there are two noteworthy experiments that have measured the time varying force on dipolar-material bodies, namely on a magnetodielectric toroid {\color{black}(James \cite{James})} and on a high-permittivity dielectric toroid {\color{black}(Walker et al.  \cite{Walker})}.  The James magnetodielectric toroid experiment verifies the Abraham macroscopic electromagnetic-field momentum as well as the Einstein-Laub electromagnetic force obtained herein, and the Walker et al. high-permittivity dielectric toroid  experiment verifies the Einstein-Laub macroscopic electromagnetic force obtained herein.  
\subsection{The Magnetodielectric Toroid}
A simplified schematic of the magnetodielectric ferrite toroid taken from James's thesis \cite{James} (with Shockley as his thesis advisor) is shown in Fig. \ref{JamesFig}.  Actually, two different ferrite toroids were used in this experiment,  one with a relative permeability of $\mu_{\rm r} \approx 16$ and a relative permittivity of $\eps_{\rm r} \approx 7$, the other with a relative permeability of $\mu_{\rm r} \approx 43$ and a relative permittivity of $\eps_{\rm r} \approx 7.6$. A current $I_0 = I \sin(\w\pm\w_0)t$ flows in the axial ($z$) direction through the center of the toroid.  A voltage $V_0 = V \sin \w t$ is applied between the metalized inner and outer circular cylindrical surfaces of the toroid.  The value of the beat frequency $\w_0/(2\pi)$ was kept at about $3$ kHz, which approximately equaled the mechanical resonant frequency of the combined toroid and piezoelectric measurement transducer that was in contact with the toroid.  The value of $\w/(2\pi)$ was varied from $10$ to $32$ kHz.  This voltage and current produce time varying magnetic fields in the azimuthal ($\phi$) direction and time varying electric fields in the radial ($r$) direction.  Because the voltage and current are time varying, each produce both a radial electric field and an azimuthal magnetic field in the toroid.  These quasielectrostatic and quasimagnetostatic fields are relatively straightforward to determine.  The voltage sets up a radial quasielectrostatic field which in turn produces an azimuthal quasimagnetostatic field that varies asymmetrically with $z$ across the toroid.  Similarly, the current sets up an azimuthal quasimagnetostatic field which in turn produces a radial quasielectrostatic field that varies asymmetrically with $z$ across the toroid.  In all, one has a total time varying radial electric field $E_r(r,z,t)$ and a total time varying  azimuthal magnetic field $H_\phi(r,z,t)$.  These electric and magnetic fields produce an axial ($z$-directed) force on the toroid that James was able to measure with a piezoelectric transducer.  The predominant measurable time-varying force occurs at the resonant beat frequency $\w_0/(2\pi)$.

\begin{figure}[ht]
\begin{center}
\scalebox{.75}
{\includegraphics[width =5.0in]{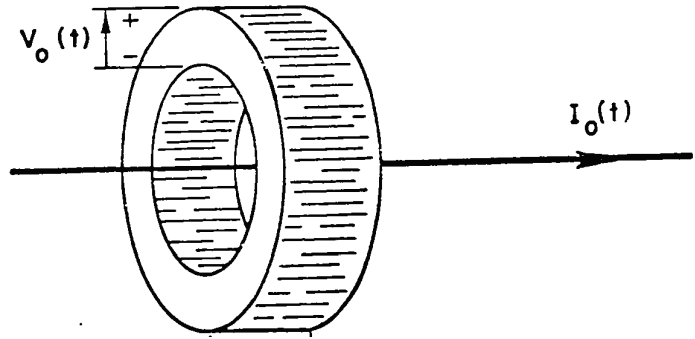}}
\end{center}
\caption*{\label{JamesFig} \small{\bf Figure 1.}  Simplified schematic of the magnetodielectric ferrite toroid used in the James experiment \cite{James}.}
\end{figure}
%
\par
The basic equation that James uses to decide experimentally between the Abraham and Minkowski electromagnetic-field momenta corresponds to our equation (\ref{agent2}), that is 
\be{James1}
\vF_{\!ag}(t) = \frac{d\vG_{\rm k}(t)}{dt} + \frac{d}{dt}\vG_f(t)   - \int\limits_S \vnh\cdot\dT dS 
\ee
where $\vF_{\!ag}(t) = F_{\!ag}(t)\vzh$ is James's transducer-measured force on the toroid.  The surface integral of the stress dyadic $\dT$ in (\ref{James1}) is evaluated analytically over the surface $S$ in free space just outside the material of the toroid, using the theoretically determined expressions for $E_r(r,z,t)$ and $H_\phi(r,z,t)$, to  give a force contribution in the axial ($z$) direction.  The kinetic-momentum force $d\vG_{\rm k}(t)/dt$ turns out to be effectively negligible at the $\w_0$ resonant beat frequency that is the dominate variation of the measured force $F_{\!ag}(t)$, so that (\ref{James1}) effectively reduces to 
\be{James2}
\vF_{\!ag}(t) = \frac{d}{dt}\vG_f(t)   - \int\limits_S \vnh\cdot\dT dS.
\ee
The $\vG_f(t)=G_f(t)\vzh$ is the electromagnetic-field momentum, which equals $\int_V\vE\times\vH/c^2 dV$ for the Abraham momentum and $\int_V\vD\times\vB dV$ for the Minkowski momentum with $V$ the volume of the toroid inside $S$.  Since $\vD = \eps_0\eps_{\rm r}\vE$ and $\vB = \mu_0\mu_{\rm r}\vH$, the Minkowski electromagnetic-field momentum equals $\mu_{\rm r}\eps_{\rm r}$ times the Abraham electromagnetic-field momentum, so that (\ref{James2}) can be rewritten as
\be{James3}
\vF_{\!ag}(t) =  \left\{\begin{array}{c} 1 \\ \mu_{\rm r}\eps_{\rm r} \\ \end{array}\right\} \frac{1}{c^2} \frac{d}{dt}\int\limits_V \vE\times\vH \,dV  - \int\limits_S \vnh\cdot\dT dS
\ee
where the $1$ in the curly bracket applies to the Abraham electromagnetic-field momentum and the $\mu_{\rm r}\eps_{\rm r}$ in the curly bracket applies to the Minkowski electromagnetic-field momentum.  The values of $\mu_{\rm r}\eps_{\rm r}$ were about $112$ and $327$ for the two different ferrite toroids that were used.
\par
James's measurements of $\vF_{\!ag}(t)$ in (\ref{James3}) along with his theoretically evaluated stress-dyadic force reveal that the Abraham electromagnetic-field momentum satisfies the equation in (\ref{James3}) much more closely than the Minkowski electromagnetic-field momentum.  This is a strong confirmation of the correctness of the Abraham electromagnetic-field momentum because both the electric and magnetic fields producing the forces and momenta are time varying and the ratio of the two momenta are equal to $112$ for one of the toroids and $327$ for the other.  Since the right-hand side of the experimentally verified (top) equation in (\ref{James3}) is equal to the negative of the Einstein-Laub force, as seen in (\ref{Mredb},\ref{4Mone}), James's experimental results also confirm the Einstein-Laub electromagnetic force as opposed to the Minkowski electromagnetic force.   Implicitly, these experimental results for magnetodielectric toroids in favor of the Abraham macroscopic electromagnetic-field momentum also confirm the existence of the internal ``hidden-momentum'' electromagnetic force within Amperian magnetic dipoles.
\subsection{The High-Permittivity Dielectric Toroid}
The paper by Walker et al. \cite{Walker} replaces the magnetodielectric toroid in Fig. \ref{JamesFig} with a high-permittivity dielectric toroid ($\eps_{\rm r} \approx 3620$, $\mu_{\rm r}= 1$), and replaces the time varying axial current $I_0(t)$ with an axial static magnetic field $\vH_0 = H_0\vzh$ penetrating the entire toroid.  The voltage $V_0(t)$ remains a sinusoidal time varying voltage applied across inner and outer conducting surfaces of the toroid.  Similarly to the magnetodielectric toroid, this voltage produces a radial quasielectrostatic field $E_{\rm r}(r,t)$ which in turn produces an azimuthal quasimagnetostatic field $H_\phi(r,z,t)$ that varies asymmetrically with $z$ across the toroid.  The azimuthal torque produced by the fields $E_{\rm r}(r,t)$, $H_\phi(r,z,t)$, and $H_0$ on the toroid was measured using a torsion pendulum arrangement.  Since the asymmetric azimuthal quasimagnetostatic field $H_\phi(r,z,t)$ exerts no azimuthal torque on the toroid, this field can be ignored and, indeed, the authors do not mention or determine this field.
\par
The theory in the Walker et al. paper begins with an expression for force density taken from Marx and Gyorgyi \cite{M&G}
\be{W1}
\frac{(\mu_{\rm r}\eps_{\rm r}-1)}{c^2}\frac{d}{dt}(\vE\times\vH) = \frac{(\eps_{\rm r}-1)}{c^2}\frac{d}{dt}[E_{\rm r}(r,t) H_0]\vphih
\ee
which is just the azimuthal ($\vphih$) component of our Einstein-Laub force density in (\ref{Mredb}) because all the force densities in the integrand of (\ref{Mredb}) are zero except for
\be{W2}
\mu_0 \frac{\partial \vP}{\partial t}\times \vH =
\frac{(\mu_{\rm r}\eps_{\rm r}-1)}{c^2}\frac{d}{dt}(\vE\times\vH) = \frac{(\eps_{\rm r}-1)}{c^2}\frac{d}{dt}[E_{\rm r}(r,t) H_0]\vphih
\ee
and {\color{black}for} $\vP\cdot\nabla\vE$, which is in the radial direction and thus can be ignored.  
\par
Essentially, Walker et al. evaluated $E_r(r,t)$ theoretically and integrated the right-hand side of (\ref{W2}) around the toroid to get the total value of the torque exerted on the toroid by the fields.  They then found that this theoretically predicted value of the torque agreed to within 10\% with the measured value of the torque.  This is a highly significant result since the corresponding value of the azimuthal torque determined by the Minkowski force in (\ref{Mredc}) is equal to zero.  Thus the experimental results of Walker et al. conclusively rule out the Minkowski macroscopic electromagnetic force while confirming the Einstein-Laub macroscopic electromagnetic force.

\section{Conclusion}
We consider a macroscopic (characterized by sources and fields spatially averaged at each instant of time over electrically small volume elements) dipolar continuum ({\color{black}a} medium obeying the Maxwell dipolar equations) with molecular dipole moments realistically modeled by classical microscopic discrete electric-charge electric dipoles and circulating-electric-current magnetic dipoles  (assuming the dipoles are densely packed enough in the defining electrically small volume elements that the macroscopic fields and polarizations can be sufficiently smoothed).   It is rigorously proven that the sum of the electromagnetic forces on a volume of these realistic classical microscopic dipoles is equal to the macroscopic force given by the Einstein-Laub formulation and, thus, the macroscopic Abraham/Einstein-Laub electromagnetic-field momentum, rather than the macroscopic Minkowski electromagnetic-field momentum (or the two other possible electromagnetic-field momenta), gives the correct electromagnetic-field momentum equal to the sum of the electromagnetic-field momentum of all the microscopic dipoles in the volume.  For periodic fields, the time derivatives of all four macroscopic electromagnetic-field momenta within the volume average to zero and all four of the associated time-averaged macroscopic forces are equal to the time-averaged total microscopic/macroscopic electromagnetic momentum flow into the volume.
\par
A key to deriving the correct macroscopic electromagnetic force on a volume of dipolar material from the sum of the individual electromagnetic forces on the discrete microscopic dipoles within the volume of material is to realize that the macroscopic polarizations, fields, and forces within a volume $V$ can be defined consistently and unambiguously if and only if the surface $S$ of the volume $V$ (like the surfaces $\Delta S$ of the averaging volume elements $\Delta V$) does not intersect the dipoles so that there are a discrete number of dipoles inside the surface $S$ (like inside the surfaces $\Delta S$). This implies that the surface $S$ of the volume $V$ in the macroscopic dipolar continuum must lie in a hypothetical thin free-space shell separating $V$ from the rest of the continuum.  With this unambiguous definition of fields and polarization densities using macroscopic volume elements surrounded by hypothetical thin free-space shells, it follows that the forces produced on the volume elements by the surface-charge and surface-current densities on either side of the thin shells cancel so that the macroscopically defined fields can be used in the force expressions rather than the cavity fields that exist in the absence of each polarized volume element.
\par
Another key to the derivation of the correct macroscopic electromagnetic force is the rigorous proof given in \cite{Yaghjian-Force} that for arbitrarily time varying externally applied electromagnetic fields, the force on a conductor model of a microscopic electric-current magnetic dipole, such as a wire loop, contains an internal-momentum  (the so-called hidden-momentum) electromagnetic force (induced indirectly by the externally applied fields) that, when added to the direct force exerted by the externally applied fields, equals the same electromagnetic force that an equal-moment, microscopic magnetic-charge magnetic dipole would experience in the same externally applied fields. (Past derivations of this ``hidden momentum'' have been confined to quasistatic rather than arbitrarily time varying electromagnetic fields and dipole moments; see \cite{Yaghjian-Force} for details.)  Moreover, after the force on the microscopic dipoles is spatially averaged in a macroscopic volume element $\Delta V$ to get the macroscopic force density, this macroscopic force density on the magnetization contains the analogous macroscopic hidden-momentum force density.
\par
With the correct macroscopic electromagnetic force determined for a volume of dipolar material in an applied external electromagnetic field, the time rate of change of the total kinetic momentum of the material in the volume can be determined through Newton's relativistic equation of motion in terms of the correct macroscopic electromagnetic force and any other forces applied to the material in the volume by an outside agent.   The difference between the kinetic  momentum and the canonical momentum (which, in classical physics, is the same as the Minkowski-force momentum) in a volume of dipolar material is shown to equal the sum of the ``hidden electromagnetic momentum''  of the circulating-electric-current magnetic dipoles and the ``hidden electromagnetic momentum'' of hypothetical circulating-magnetic-current electric dipoles replacing the electric-charge electric dipoles within the volume of material.  These electric-current magnetic dipoles and magnetic-current electric dipoles that exhibit hidden momentum are required in the Minkowski $(\vD,\vB)$ formulation of Maxwell's equations and the constitutive relations, as opposed to the Abraham $(\vE,\vH)$ formulation that requires electric- and magnetic-charge dipoles, which exhibit no hidden momentum.
\par
Most experiments in the past that measured radiation forces in media revealed only the time-averaged electromagnetic forces for which there is no difference between the Einstein-Laub/Abraham and Minkowski formulations.  However, two well-conceived and well-conducted past experiments that measured the time varying macroscopic electromagnetic forces and momenta are shown to decidedly confirm the force and momentum expressions of the Einstein-Laub/Abraham formulation and to rule out those of the Minkowski formulation.
\ack
This work was supported under the U.S. Air Force Office of Scientific Research Contract \# FA9550-22-1-0293 through Dr. Arje Nachman.
%
\renewcommand\refname{\large References} 

}
\end{document}